\documentstyle[12pt]{article}
\newcommand{\eq}{\begin{equation}}
\newcommand{\en}{\end{equation}}
\newcommand{\eqn}{\begin{eqnarray}}
\newcommand{\enn}{\end{eqnarray}}
\newcommand{\CR}{\nonumber \\}
\newcommand{\sg}{\sigma}
\newcommand{\th}{\theta}
\newcommand{\Th}{\Theta}
\newcommand{\pa}{\partial}
\newcommand{\A}{\alpha}
\newcommand{\Ap}{\alpha^{\pr}}

\newcommand{\ep}{\epsilon}
\newcommand{\no}{\nonumber}

\newcommand{\fr}{\frac}
\newcommand{\pr}{\prime}

\newcommand{\ket}{\rangle}
\newcommand{\EQ}{\begin{equation}}
\newcommand{\EN}{\end{equation}}
\newcommand{\bea}{\begin{eqnarray}}
\newcommand{\ena}{\end{eqnarray}}
\newcommand{\vs}[1]{\vspace{#1 mm}}
\newcommand{\hs}[1]{\hspace{#1 mm}}

\renewcommand{\d}{\delta}
\newcommand{\e}{\epsilon}
\newcommand{\nn}{\nonumber\\}
\newcommand{\p}[1]{(\ref{#1})}

\begin{document}
\renewcommand{\thefootnote}{\fnsymbol{footnote}}
\begin{titlepage}
\begin{flushright}
hep-th/9801135   \\
UT-Komaba/98-1 \\
OU-HET 289 \\
\end{flushright}
\vspace{0.5cm}
\begin{center}
{\Large \bf
Fermionic Zero Mode and String Creation between D4-Branes at Angles
}
\vs{15}

{\large
Takuhiro Kitao\footnote{e-mail address: kitao@hep1.c.u-tokyo.ac.jp}} \\
\vs{5}
{\em Institute of Physics, University of Tokyo,
Komaba, Tokyo 153, Japan}\\
\vs{5}
{\large
and \\
\vs{5}
Nobuyoshi Ohta\footnote{e-mail address: ohta@phys.wani.osaka-u.ac.jp}
and
Jian-Ge Zhou\footnote{e-mail address: jgzhou@phys.wani.osaka-u.ac.jp,
JSPS postdoctral fellow}} \\
\vs{5}
{\em Department of Physics, Osaka University,
Toyonaka, Osaka 560, Japan}

\end{center}
\vs{15}
\centerline{{\bf{Abstract}}}
\vs{5}

We study the creation of a fundamental string between D4-branes at
angles in string theory. It is shown that ${\rm R}(-1)^{\rm F}$
part of the one-loop potential of open string changes its sign due
to the change of fermionic zero-mode vacua when the branes cross each other.
As a result the effective potential is independent of the angles when
supersymmetry is partially unbroken, and leads to a consistent picture
that a fundamental string is created in the process.
We also discuss the s-rule in the configuration.
The same result is obtained from the one-loop potential for the orthogonal
D4-branes with non-zero field strength.
The result is also confirmed from the tension obtained by deforming the
Chern-Simons term on one D4-brane, which is induced by another tilted D4-brane.
\end{titlepage}
\renewcommand{\thefootnote}{\arabic{footnote}}
\setcounter{footnote}{0}
\baselineskip=0.7cm

Recently it has been pointed out from various points of view that
a brane (string) is created when certain two branes cross each other
\cite{HW}-\cite{BG}. This is originally suggested in the field 
theory analysis on D-branes in ref.~\cite{HW}. Consistencies of this result
are confirmed by deforming the Chern-Simons term of the system~\cite{DFK}
and by the one-loop potentials of open strings~\cite{BGL}. Moreover,
the results of M(atrix) theory also support this fact~\cite{HWu,OSZ} and
the authors of refs.~\cite{NOYY,Y,BGS} have explained the brane creation by
using the technique of holomorphic embedding of M5-brane or M2-brane
into Taub-NUT space induced by Kaluza-Klein monopole.

In this paper, we study the creation of a fundamental string between two
D4-branes at angles, in which the configuration has less supercharges than
those of orthogonal case \cite{BDL}-\cite{OT}. This system was examined in
the framework of the M(atrix) theory in ref.~\cite{OSZ}, and we intend to
study it and its variant in string theory in more detail. Of course,
we expect that the result is the same as that of M(atrix) theory~\cite{OSZ}
because the approach of M(atrix) theory necessarily involves the theory of
bound states of D-branes whose T-duality is the theory of branes with angles
discussed in ref.~\cite{OZ}. However, we find that there is a subtle issue
which needs clarification.

To discuss the string creation, we use the
one-loop potential due to the open strings between the branes at angles.
Amplitudes for such system have been obtained in ref.~\cite{J1} to
determine the condition of unbroken supersymmetry, and we can simply
read off the potential from that work. We also obtain 
the one-loop potential from open strings between the orthogonal branes
with condensation which is similar to the case with angles. When
supersymmetry is partially unbroken, these potentials vanish.
The problem is what happens when one brane adiabatically crosses
another brane, keeping the partially unbroken supersymmetry. The
authors of ref.~\cite{BGL} have interpreted this situation as the
system of brane-anti-brane, and obtained the result which is
consistent with the creation of a fundamental string, that is, the effective
potential of the string tension times the distance between the branes, which
is canceled by a string created between these branes. They discussed the case
that D0-brane passes through D8-brane, in which the ${\rm NS}(-1)^{\rm F}$
term vanishes because of the fermionic zero modes.

On the other hand, in the case with angles, ${\rm NS}(-1)^{\rm F}$ term
gives generally non-zero contribution which depends on angles.
As a consequence, it turns out that if we interpreted the system as
that of brane-anti-brane when one brane adiabatically crossed another, the
potential would be dependent on the angles. This is a very strange result
if we try to understand it in terms of a string stretched between the two
branes. We find that in our present situation only the ${\rm R}(-1)^{\rm F}$
term changes the sign after crossing because the vacua defined by fermionic
zero mode change. The result is then independent of the angles and is given by
the string tension times the distance between the branes, allowing the physical
interpretation in terms of string creation. The potential is canceled by the
contribution from the string created between the branes.

In the case of orthogonal two D-branes, it has been pointed out in
ref.~\cite{DFK} that the force induced by NS and R terms which is one
half of the string tension~\cite{L1} cancels the tension induced
by Chern-Simons term. We also calculate this tension from the Chern-Simons
term in the case with angles. We find that the result is independent of
the angles and is equal to one half of string tension, the same as that
found in ref.~\cite{DFK}. This confirms our result that the effective
potential is independent of the angles and only the ${\rm R}(-1)^{\rm F}$
term changes the sign after brane crossing.

Our results also indicate that the s-rule of Hanany and Witten is
valid~\cite{HW}. The rule translated into our present setting states
that a configuration with more than one fundamental string joining the two
D4-branes cannot be supersymmetric. As we will see, this rule can be
understood as the uniqueness of the chiral fermionic zero-mode vacuum.

Let us start with the configuration of two D4-branes, one of which has tilted
world-volume to that of another D4-brane. The boundary conditions of the open
string on one D4-brane (denoted as D4) are
\eqn
&& \pa_{\sg} X^{\mu} = 0, \ \ \ \ \mu =0,\ldots ,4, \CR
&& X^{\mu} = 0, \ \ \ \ \mu =5,\ldots ,9,
\label{1}
\enn
at $\sg =0$, where $\sg$ is the world-sheet coordinate which
spans $[0, \pi]$.  The boundary conditions
for fermions follow from the world-sheet supersymmetry
\bea
\d X^\mu = {\bar \e} \psi^\mu.
\label{super}
\ena
Those at $\sg =\pi$ on another D4-brane (denoted as D$4^{\pr}$) are
\eqn
&&\pa_{\sg} X^{0} = 0, \CR
&& \pa_{\sg} X^{i} \cos (\th_i \pi) - \pa_{\sg} X^{i+4} \sin
 (\th_i \pi) = 0, \CR
&& X^{i} \sin (\th_i \pi) + X^{i+4} \cos (\th_{i} \pi) = 0,
 \ \ \ \ i=1,\ldots ,4, \CR
&& X^{9} = b,
\enn
where $b$ is the distance between the two branes and $\{{\th}_i \pi \}$
are the angles parameterizing our two D4-brane configuration.

The one-loop amplitude of the open string which satisfies the above
boundary conditions has been obtained in ref.~\cite{J1}. Assuming that
all the four angles are non-zero,\footnote{For zero angle $\th_i$, it
is easy to do a similar calculation. One finds that the summation of
the bosonic degrees of freedom is replaced by the product of the integral
over the momentum $p_i$ and the volume of the direction $X^i$. It should
be possible to continuously interpolate the zero angle limit in eq.~\p{pot1},
but it seems complicated to do so explicitly.}
we find that the potential between these D4-branes is
\bea
V &=& - \int_{0}^{\infty} \fr{dt}{2t} \fr{{\rm
e}^{-\fr{b^2 t}{2 \pi \A^{\pr}}}}{(8 {\pi}^2 \A^{\pr} t)^{\fr{1}{2}}}
\left[ - \prod_{k=1}^{4} \fr{\Th_{2}(i \th_k t | it) }{\Th_{1}(i \th_k
t | it)} + \prod_{k=1}^{4} \fr{\Th_{3}(i \th_k t | it) } {\Th_{1}(i
\th_k t | it)} \right. \nn
&& \hs{20}\left. + \prod_{k=1}^{4} \fr{\Th_{1}(i \th_k t | it) }
{\Th_{1}(i \th_k t | it)} - \prod_{k=1}^{4} \fr{\Th_{4}(i \th_k t |
it) }{\Th_{1}(i \th_k t | it)} \right].
\label{pot1}
\ena
The terms in the bracket are the contributions from R, NS, R$(-1)^{\rm F}$
and NS$(-1)^{\rm F}$ sectors, respectively. The author of ref.~\cite{J1} has
derived the condition that the amplitude vanishes, in the search for the
criterion that supersymmetry is partially unbroken. It is given by
\eq
\th_1 \pm \th_2 \pm \th_3 \pm
\th_4 =0, \ \ \ \ \ {\rm mod} \ \ 2,
\label{cond}
\en
for which we have
\EQ
\prod_{k=1}^{4} \Th_{1}(i \th_k t|it)
 - \prod_{k=1}^{4} \Th_{2}(i \th_k t|it)
 + \prod_{k=1}^{4} \Th_{3}(i \th_k t|it)
 - \prod_{k=1}^{4} \Th_{4}(i \th_k t|it) = 0.
\label{cond1}
\EN
Indeed, eq.~\p{cond} is the same as that derived in supergravity in
ref.~\cite{OT}, in which it has been shown that there is
1/16 unbroken supersymmetry if eq.~\p{cond} is satisfied.
In what follows, we will mainly concentrate on the case \p{cond}.

Let us consider what happens when the above one brane passes through another
adiabatically in the direction of $x^9$ from $x^9=|b|$ to $x^9=-|b|$.
We can regard this situation as the sum of two
parts. One is that of one of the branes hooked by another brane, which
is expected to be a fundamental string.\footnote{More concrete picture
of this ``hooking'' will be explained below.} Another part is that from
the two branes passing by without hooking. If the interaction between
the two branes of the latter part in this situation is the string
tension times the distance between the branes, the first part is
expected to represent the created fundamental string to cancel this
interaction between the two branes.

What is the difference between the state before crossing (``configuration
[A]'') in which 1/16 SUSY is unbroken and that after crossing
(``configuration [B]'') ignoring the hooking part?
The only difference is the vacua defined by the
R-fermion zero modes, which exist only in the world-sheet fermions
$\psi^0$ and $\psi^9$. There are no other zero modes
because the boundary conditions with the angles shift the R-fermion
zero modes of the rotated directions. We define the vacua $| \pm \ket$ by
\eq
({\psi}^0_0 \pm {\psi}^9_0 )| \pm \ket =0.
\label{vac} 
\en 
By GSO projection, either $+$ or $-$ is projected out, so that
only one space-time massless chiral fermion can exist when the two
branes intersect. Configuration [A] is related to [B] by the parity
transformation in the direction of $x^9$. In order to preserve the
supersymmetry \p{super}, $\psi^9_0$ transforms into $-\psi^9_0$ when
the configuration changes from [A] to [B]. It follows that the vacuum of
[A] is different from that of [B]. In other words, the definition of GSO
projection for the R-sector is different between [A] and [B]. As a
result, the potential of [B] is different from [A] by the sign of the
${\rm R}(-1)^{\rm F}$ term. By using the condition of unbroken
supersymmetry~\p{cond1}, we find
\eqn
V &=& - \fr{1}{2} \int_{0}^{\infty} \fr{dt}{t}
 \fr{{\rm e}^{-\fr{b^2 t}{2 \pi\A^{\pr}}}}{(8 \pi^2 \A^{\pr} t)^{\fr{1}{2}}}
 \left[- \prod_{k=1}^{4}\fr{\Th_{2}(i \th_k t | it) }{\Th_{1}(i \th_k t | it)}
 + \prod_{k=1}^{4} \fr{\Th_{3}(i \th_k t | it) }{\Th_{1}(i \th_k t | it)}
 \right.\nn
&&\left.\hs{20} \pm \prod_{k=1}^{4}
 \fr{\Th_{1}(i \th_k t |it)}{\Th_{1} (i \th_k t | it)}
- \prod_{k=1}^{4} \fr{\Th_{4}(i \th_k t|it)}{\Th_{1}(i \th_k t|it)} \right] \CR
&=& - \fr{1}{2}
\int_{0}^{\infty} \fr{dt}{t} \fr{{\rm e}^{-\fr{b^2 t}{2 \pi
\A^{\pr}}}}{(8 \pi^2 \A^{\pr} t)^{\fr{1}{2}}} (-1 \pm 1) \nn &=& -
\fr{T_0}{2} ( 1 \mp 1) \mid b \mid ,
\label{tension}
\enn
where $-(+)$ in the last equation is for the configuration [A] ([B]),
and $T_0$ is the string tension $\fr{1}{2\pi \A^{\pr}}$.
This change of the sign occurs when the two branes intersect
each other at $b=0$.
We see that the potential computed in eq.~\p{tension} for [B] is
independent of the angles and equals the string tension times the
distance. On the other hand, since we move one brane across another from
$x^9=|b|$ to $x^9=-|b|$ adiabatically, the fermionic zero-mode
vacuum is changed after crossing from $x^9=0$ all the way down to
$x^9=-|b|$ and we get contributions from that part
which exactly cancel the above potential. This is what we have called
``hooking'' in the above. Physically this should be understood as due
to a string created between the two branes when supersymmetry is
unbroken. In this way, the BPS property of the system is preserved
before and after the brane crossing.

It is crucial that only the R$(-1)^{\rm F}$ term changes its sign;
if we interpreted the final configuration as brane-anti-brane system,
both NS$(-1)^{\rm F}$ and R$(-1)^{\rm F}$ terms (related by modular
transformation to the RR sector in the
closed string channel) would change the signs, resulting in a potential
dependent on the angles. Such potential could not be canceled by string
tension, and would allow no physical interpretation.\footnote{In the absence
of NS$(-1)^{\rm F}$ term, these pictures make no difference, and both
give the same consistent results for string creation~\cite{BGL}.}

The fact that there is only one space-time massless chiral fermion exists
at the intersection is closely related to the
s-rule~\cite{HW,BGS,BG}. Our results indicate that only a single
string is stretched between two D4-branes after crossing.  By a chain
of dualities, the special orthogonal case can be transformed to the
configuration in which a D3-brane is suspended between an NS 5-brane and
a D5-brane considered in ref.~\cite{HW}. Thus our results give the
generalization of this rule to more general angles.

It is easy to repeat the calculation when any one of the angles $\th_i$
is zero under the condition (\ref{cond}), and we find that the potential
vanishes and no string is created.

We can derive the same result in the case of orthogonal D4-D4$'$ system
with bound state~\cite{ACNY}-\cite{J2}. The boundary conditions of
the open string on one D4-brane are
\eqn
&& \pa_{\sg} X^{0} = 0, \CR && \pa_{\sg} X^{2k-1} + 2
\pi \Ap F_{(k)} \pa_{\tau} X^{2k} = 0, \CR
&& \pa_{\sg} X^{2k} -
2 \pi \Ap F_{(k)} \pa_{\tau} X^{2k-1} = 0,\ \ \ k=1,\ \ 2, \CR
&& X^{\mu} = 0, \ \ \ \mu=5, \ldots 9,
\enn
at $\sg =0$, where $\tau$
is the world-sheet coordinate along with $\sg$ and $F_{(k)}$ is
the condensation of the field strength. The boundary conditions at
$\sg =\pi$ on another D4$'$-brane are
\eqn 
&& \pa_{\sg} X^{0} = 0, \CR 
&& X^{\mu} = 0, \ \ \ \ \mu=1, \dots ,4, \CR && \pa_{\sg} X^{2k-1} + 2 \pi
\Ap F_{(k)} \pa_{\tau} X^{2k} = 0, \CR && \pa_{\sg} X^{2k} - 2
\pi \Ap F_{(k)} \pa_{\tau} X^{2k-1} = 0,\ \ \ k=3,\ \ 4,\CR &&
X^{9} = b.
\enn
By repeating the calculation similar to the above, we obtain the potential
\bea
&& V = - \int_{0}^{\infty} \fr{dt}{2t} \fr{{\rm e}^{-\fr{b^2 t}{2 \pi\A^{\pr}}}}
{(8 \pi^2 \A^{\pr} t)^{\fr{1}{2}}} \left[ - \prod_{k=1}^{4}
\fr{\Th_{3}(i \ep_k t | it) }{\Th_{4}(i \ep_k t |it)} +
\prod_{k=1}^{4} \fr{\Th_{2}(i \ep_k t|it)}{\Th_{4}(i \ep_k t|it)} \right.\nn
&& \hs{30} \left. \pm \prod_{k=1}^{4} \fr{\Th_{4}(i \ep_k t|it)}
{\Th_{4}(i \ep_k t|it)}
- \prod_{k=1}^{4} \fr{\Th_{1}(i \ep_k t|it)}{\Th_{4}(i \ep_k t|it)} \right],
\label{pot2}
\ena
where $\ep_k \equiv \fr{{\tan}^{-1}(2 \pi \Ap F_{(k)})}{\pi}$ and $\pm$
corresponds to the difference in the definitions of GSO projection.
The condition for 1/16 unbroken SUSY is the same as the case with angles:
\eq
\ep_1 \pm \ep_2 \pm \ep_3 \pm
\ep_4 =0, \ \ \ \ \ {\rm mod} \ \ 2. \no
\en
With the help of eq.~\p{cond1}, the potential~\p{pot2} is cast into
\eq
V= - \int_{0}^{\infty} \fr{dt}{2t} \fr{{\rm e}^{-\fr{b^2 t}{2 \pi
\A^{\pr}}}}{(8 \pi^2 \A^{\pr} t)^{\fr{1}{2}}} ( -1 \pm 1)
= -\fr{T_0}{2} ( 1 \mp 1) \mid b \mid .
\label{tensioncond}
\en
This also shows that a fundamental string is created.

As a further check of our results, let us calculate the induced tension from
the Chern-Simons term. We again deal with the system of two D4-branes
at angles. Following ref.~\cite{DFK} in which the system of D0 and D8-branes
is discussed, we consider the Chern-Simons term on the first D4:
\eq
\fr{\mu_{{\rm D4}}}{4!} \int d^5 x
\ep_{{\nu_0}{\nu_1}{\nu_2}{\nu_3}{\nu_4}}
F_{(4)}^{{\nu_0}{\nu_1}{\nu_2}{\nu_3}} A^{{\nu_4}},
\label{cs}
\en
where $F_{(4)} \equiv d C_{(3)}$, $C_{(3)}$ is the R-R 3-form gauge
field, and $A^{\nu}$ is the U(1) gauge field on D4. The indices $\{
{\nu_k} \}$ are those of the coordinates of the world-volume of the first D4,
$x^{0,1, \ldots ,4}$. The second D4$'$-brane is the source of
$F_{(4)}= *F_{(6)}$ in the integral~\p{cs}.
Assuming that $F_{(4)}$ depends on only $x^1, \ldots ,x^4$ and that
$A^{\nu}$ depends on only $x^0$, the above term reduces to
\eq
\mu_{ {\rm D4} } \int(\prod_{i=1}^4 dx_i ) F_{(4)}^{1234}
\int ds \fr{dx_0}{ds} A^{0}(x_0), \no
\label{ten}
\en
where $s$ is a parameter on the D4 and can be supposed to be a world-sheet
coordinate, and then $ \mu_{{\rm D4}} \int(\prod_{i=1}^4 dx_i) F_{(4)}^{1234}$
is interpreted as the tension to be evaluated below.

On the other hand, the R-R charge $\mu_{{\rm D4}^{\pr}}$ of
${\rm D4}^{\pr}$ is
\eq
\mu_{{\rm D4}^{\pr}} = \int_{S_4} *F_{(6)}
 d^4 x = \int_{S_4} F_{(4)} d^4 x, \no
\en
where $S_4$ is the 4-sphere surrounding ${\rm D4}^{\pr}$. Hence
${\rm F}_{(4)}$ can be taken as
\eq
\fr{1}{4!}\ep_{{\nu_0^{\pr}}{\nu_1^{\pr}}{\nu_2^{\pr}}{\nu_3^{\pr}}
{\nu_4^{\pr}}}
F_{(4)}^{{\nu_0^{\pr}}{\nu_1^{\pr}}{\nu_2^{\pr} }{\nu_3^{\pr}} }
 = \fr{{\mu}_{{\rm D4}^{\pr}}}{r^4 {\Omega}_4}
 \fr{x_{{\nu}_4^{\pr}}}{r},
\label{FFF} 
\en 
where all $\nu^{\pr}$s are the indices of the directions which are
orthogonal to the world-volume of D4$'$, and $r$ and $\Omega_4$
are the radius of $S_4$ and the volume of a unit 4-sphere, respectively.
We denote the directions of the world-volume of D4$'$ as
$x_0^{\pr}, \ldots ,x_4'$. Then the non-zero components of
$F_{(4)}^{{\nu_0'}{\nu_1'}{\nu_2'}{\nu_3'}}$ are those with indices
outside these dimensions:
\eq
F_{(4)}^{5^{\pr} 6^{\pr} 7^{\pr} 8^{\pr}},\ \
F_{(4)}^{6^{\pr} 7^{\pr} 8^{\pr} 9^{\pr}},\ \
F_{(4)}^{7^{\pr} 8^{\pr} 9^{\pr} 5^{\pr}},\ \
F_{(4)}^{8^{\pr} 9^{\pr} 5^{\pr} 6^{\pr}},\ \
F_{(4)}^{9^{\pr} 5^{\pr} 6^{\pr} 7^{\pr}}. \no
\en

To evaluate the contribution to $F_{(4)}^{1234}$ at a point Q on D4 from
a point P on the second ${\rm D4}^{\pr}$, let us consider
the 5-dimensional plane which intersects ${\rm D4}^{\pr}$ orthogonally
at P, and also D4 at Q. The relation between the coordinates $\{ x_i\}$
for D4 and $\{x_i'\}$ for D4$'$ is
\eqn
&& x_i = x_i^{\pr} {\rm cos} ({\th}_i \pi ) + x_{i+4}^{\pr} {\rm sin}
( \th_i \pi), \CR
&& x_{i+4} = x_{i+4}^{\pr} {\rm cos} ( \th_i \pi )
 - x_{i}^{\pr} {\rm sin} (\th_i \pi), \ \ \ \ i=1,\ldots ,4, \CR
&& x_{9} = x_{9}^{\pr} + b.
\enn
We see that the coordinates $x'_{5,\cdots,8}$ mix with $x_{1,\cdots,4}$ with
the coefficients $\sin(\th_i \pi)$ and hence $F^{1234}$ in \p{ten} gets
the following contribution from $F^{5'6'7'8'}$:
\eq
F_{(4)}^{1234}
= \Big( \prod_{i=1}^{4} \sin (\th_i \pi ) \Big)
 F_{(4)}^{5^{\pr}6^{\pr}7^{\pr}8^{\pr}} =  - \fr{{\mu}_{{\rm
D4}^{\pr}}}{r^4 \Omega_4} \fr{b}{r}
\Big( \prod_{i=1}^{4} \sin ({\th}_i \pi ) \Big),
\label{FF} 
\en
where $r$ is the distance between $P$ and $Q$, and we have used eq.~\p{FFF}
and the fact that $P$ is away from $Q$ by $b$ in the direction of $x^9$.
Let us take the origin on D4 at the nearest point from D4$'$ and
denote the coordinates of Q as $(y_1,y_2,y_3,y_4)$. Then
$r=$PQ is given by $r^2 = b^2 + \sum_{i=1}^{4}y_i^2 {\rm sin}^2 (\th_i
\pi )$. Using this expression and eq.~(\ref{FF}), we obtain
\eqn
&&\mu_{ {\rm D4} } \int
(\prod_{i=1}^4 dy_i ) F_{(4)}^{1234} = \mu_{ {\rm D4} } \int
(\prod_{i=1}^4 dy_i ) \left( \prod_{i=1}^{4} \sin ( \th_i \pi) \right)
F_{(4)}^{5^{\pr} 6^{\pr} 7^{\pr} 8^{\pr}} \CR
&& =  - \mu_{\rm D4}
\int (\prod_{i=1}^4 dy_i ) \left( \prod_{i=1}^{4} \sin (\th_i \pi)\right)
\fr{\mu_{{\rm D4}^{\pr}}}{r^4 \Omega_4} \fr{b}{r}
= - \fr{1}{2} \mu_{ {\rm D4} } \mu_{{\rm D4}^{\pr}} \fr{b}{ \mid b \mid },
\label{tenr}
\enn
which is independent of the angles.
Because $\mu_{ {\rm D4} }= \mu_{{\rm D4}^{\pr}}= (\fr{1}{{2 \pi
\A^{\pr}}})^{\fr{1}{2} }$, eq.~\p{tenr} turns out to be
$- \fr{1}{4 \pi \A^{\pr}}\fr{b}{ \mid b \mid } = - \fr{T_0}{2}
\fr{b}{ \mid b \mid }$, that is, one half contribution
of the interaction between a fundamental string and a
D-brane.\footnote{The overall sign is not significant.}
This result is the same as the orthogonal case of \cite{DFK}.

Let us compare this result with our previous discussions on the potential.
Consider the configuration [A]. The above tension is the same as the force from
the R$(-1)^F$ term in (\ref{pot1}) up to sign even if (\ref{cond}) is
not satisfied. This means that the anomaly term corresponds to
the ${\rm R}(-1)^{\rm F}$ term in the potential~\p{pot1}.\footnote{When one
of the four angles is zero, both terms vanish, irrespectively of whether
supersymmetry is unbroken or not.}
Another half contribution is expected to appear so that
the force is canceled because of unbroken supersymmetry~\cite{DFK,L1}.
This term may come from the effective action when we
integrate the fermion with the lightest mass in supergravity.
In fact the authors of \cite{OSZ} have derived the same term by doing so
in M(atrix) theory.

To consider the configuration [B], we divide the one-loop potential of
configuration [A] into two pieces as
\eq
{\rm R} + {\rm NS} + {\rm NS}(-1)^F + {\rm R}(-1)^F
= \Big( {\rm R} + {\rm NS} + {\rm NS}(-1)^F - {\rm R}(-1)^F \Big)
+ 2 {\rm R}(-1)^F.
\en
The left hand side is the potential of configuration [A] which vanishes by
supersymmetry. The first term on the right hand side is the potential for [B]
which we computed in eq.~\p{tension} and the second term is for the created
string. The double of the force induced by the anomaly term corresponds to
that of the created string, which gives the same picture as in ref.~\cite{DFK}.

String creation in type I$'$ theory has also been discussed in
refs.~\cite{BGL,BGL2,BGS}. It should be straightforward to generalize our
above discussions to similar configurations in such theory, and this
will not be discussed in this paper.

In summary,
we have derived the results which are consistent with the creation of
a fundamental string in the crossing process of D4-branes at angles and its
variant configuration by using the potentials of string and deformation of
the anomaly term. It is consistent with the results of M(atrix)
theory~\cite{OSZ}. In the process, we have clarified how the configurations,
and in particular, the fermionic zero-mode vacua change, giving a consistent
picture of string creation. Our results also confirm the s-rule discussed
for orthogonal case.

\section*{Acknowledgments}
T.K. would like to thank the people of Particle Theory Group at Komaba
for helpful comments, especially T. Yoneya for useful
discussions. The work of N.O. and J.-G.Z. was supported in part by
Grand-in-aid from the Ministry of Education, Science, Sports and
Culture No. 96208.

\newcommand{\NP}[1]{Nucl.\ Phys.\ {\bf #1}}
\newcommand{\AP}[1]{Ann.\ Phys.\ {\bf #1}}
\newcommand{\PL}[1]{Phys.\ Lett.\ {\bf #1}}
\newcommand{\NC}[1]{Nuovo Cimento {\bf #1}}
\newcommand{\CMP}[1]{Comm.\ Math.\ Phys.\ {\bf #1}}
\newcommand{\PR}[1]{Phys.\ Rev.\ {\bf #1}}
\newcommand{\PRL}[1]{Phys.\ Rev.\ Lett.\ {\bf #1}}
\newcommand{\PRE}[1]{Phys.\ Rep.\ {\bf #1}}
\newcommand{\PTP}[1]{Prog.\ Theor.\ Phys.\ {\bf #1}}
\newcommand{\PTPS}[1]{Prog.\ Theor.\ Phys.\ Suppl.\ {\bf #1}}
\newcommand{\MPL}[1]{Mod.\ Phys.\ Lett.\ {\bf #1}}
\newcommand{\IJMP}[1]{Int.\ Jour.\ Mod.\ Phys.\ {\bf #1}}
\newcommand{\JP}[1]{Jour.\ Phys.\ {\bf #1}}


\end{document}